\documentstyle[epsfig,12pt,preprint,tighten,aps]{revtex}
\begin{document}

\draft

\title{\rightline{{\tt July 2000}}
\rightline{{\tt UM-P-030/2000}} \rightline{{\tt RCHEP-006/2000}} \ \\
Further studies on relic neutrino asymmetry generation II: \\a
rigorous treatment of repopulation in the adiabatic limit}
\author{Keith S. M. Lee, Raymond R. Volkas and Yvonne Y. Y. Wong}
\address{School of Physics\\
Research Centre for High Energy Physics\\ The University of
Melbourne Vic 3010\\ Australia\\ (keith.lee, r.volkas,
ywong@physics.unimelb.edu.au)}

\maketitle

\begin{abstract}
We derive an approximate relic neutrino asymmetry evolution
equation that systematically incorporates repopulation processes
from the full quantum kinetic equations (QKEs).  It is shown that
in the collision dominant epoch, the said equation reduces
precisely to the expression obtained previously from the
static/adiabatic approximation. The present treatment thus
provides a rigorous justification for the
seemingly incongruous assumptions of a negligible repopulation
function and instantaneous repopulation sometimes employed in
earlier works.
\end{abstract}

\section{Introduction}

The computationally convenient and rewarding static approximation
\cite{longpaper} may be rigorously derived from the quantum
kinetic equations (QKEs) \cite{qke,qke2} in the adiabatic
Boltzmann limit \cite{bvw,v&w}, assuming that the repopulation or
refilling function is insignificantly small.  In actual numerical
studies, however, we also adopt the view that a depleted neutrino
momentum state is instantaneously refilled from the background
plasma.  The latter condition naturally implies that while the
repopulation function may be small, it cannot be identically zero.

In principle, simultaneous suppositions of a vanishing refilling
function and instantaneous repopulation are somewhat
contradictory, or at least uncontrolled.  However, their combined
validity in the collision dominant epoch is strongly supported by
numerical evidence \cite{longpaper,v&w}.  Indeed, instantaneous
repopulation {\it without} the assumption of a negligible
repopulation function can be consistently implemented as a
sensible approximation to the QKEs \cite{p&r}.

Commencing with the full-fledged QKEs, we demonstrate rigorously
that the two aforementioned approximations,
conceived originally for computational convenience, are indeed
appropriate in a certain limit.  Our approach differs from the
fully consistent instantaneous repopulation approximation
introduced in Ref.\ \cite{p&r},
since we do not make any {\it a priori}
assumptions regarding the repopulation rate, instantaneous or
otherwise. Instead, the present work shows that a ``brute force''
treatment of the QKEs incorporating a finite repopulation rate
does indeed lead to approximate evolution equations that reduce to
those following from the dual approximations of a negligible
refilling function and instantaneous repopulation
 in the temperature regime of interest.

\section{Nomenclature}

Consider a two-flavour system comprising an active species
$\nu_{\alpha}$ (where $\alpha = e$, $\mu$ or $\tau$), and a
sterile species $\nu_s$, whose properties are characterised by the
 density matrix \cite{qke,qke2}
\begin{equation}
\rho(p) = \frac{1}{2} [ P_0(p) + {\bf P}(p) \cdot \sigma ],
\end{equation}
in which the variables $P_0$ and ${\bf P}(p) = P_x(p) \hat{x} +
P_y(p) \hat{y} + P_z(p) \hat{z}$ are functions of both time $t$
and momentum $p$, and $\sigma = \sigma_x \hat{x} + \sigma_y
\hat{y} + \sigma_z \hat{z}$ are the Pauli matrices. The diagonal
entries of $\rho$ represent respectively the $\nu_{\alpha}$ and
$\nu_s$ distribution functions in momentum space, that is,
\begin{eqnarray}
N_{\alpha}(p) &=& \frac{1}{2} [P_0(p) + P_z(p)]N^{\text{eq}}(p,0),
\nonumber \\ N_s(p) &=&\frac{1}{2} [P_0(p) -
P_z(p)]N^{\text{eq}}(p,0),
\end{eqnarray}
where the reference distribution $N^{\text{eq}}(p,0)$ is defined
as the equilibrium Fermi--Dirac function,
\begin{equation}
N^{\text{eq}}(p,\mu) = \frac{1}{2 \pi^2}
\frac{p^2}{1+\exp\left(\frac{p-\mu}{T}\right)},
\end{equation}
with chemical potential $\mu=0$ at temperature $T$.

The evolution of $P_0(p)$ and ${\bf P}(p)$ is governed by the
quantum kinetic equations (QKEs)
\begin{eqnarray}
\label{qkes} \frac{\partial {\bf P}}{\partial t} &=& {\bf V}(p)
\times {\bf P}(p) - D(p)[P_x(p) \hat{x} + P_y(p) \hat{y}] +
\frac{\partial P_0}{\partial t} \hat{z},\nonumber
\\ \frac{\partial P_0}{\partial t} &=& R_{\alpha}(p).
\end{eqnarray}
Here, the quantity ${\bf V}(p) = \beta(p) \hat{x} + \lambda(p)
\hat{z}$ is the matter potential vector \cite{rn}, with
\begin{eqnarray}
\label{betalambda} \beta (p) &=& \frac{\Delta m^2}{2p} \sin 2
\theta_0, \nonumber\\ \lambda (p) &=& \frac{\Delta m^2}{2p}[b(p) -
a(p) - \cos 2 \theta_0],
\end{eqnarray}
in which $\Delta m^2$ is the mass-squared difference between the
neutrino states, $\theta_0$ is the vacuum mixing angle, and
\begin{eqnarray}
\label{ap} a(p) &=& -\frac{4 \zeta(3) \sqrt{2} G_F L^{(\alpha)}
T^3 p}{\pi^2 \Delta m^2}, \nonumber \\ b(p) &=& -\frac{4 \zeta(3)
\sqrt{2} G_F A_{\alpha} T^4 p^2}{\pi^2 \Delta m^2 m^2_W},
\end{eqnarray}
given that $G_F$ is the Fermi constant, $m_W$ the $W$-boson mass,
$\zeta$ the Riemann zeta function and $A_e \simeq 17$,
$A_{\mu,\,\tau} \simeq 4.9$.  The effective total lepton number
(for the $\alpha$-neutrino species)
\begin{equation}
\label{efflep} L^{(\alpha)} = L_{\alpha}+L_e + L_{\mu} + L_{\tau}
+ \eta \equiv 2L_{\alpha} + \tilde{L},
\end{equation}
combines all asymmetries individually defined as $L_{\alpha} =
(n_{\nu_{\alpha}} - n_{\overline{\nu}_\alpha})/n_{\gamma}$, 
where the symbol
$n$ denotes number density, and $\eta$ is a small term due to the
cosmological baryon asymmetry.

The decoherence function $D(p)$ is related to the total collision
rate for $\nu_{\alpha}$, $\Gamma(p)$, via \cite{qke}
\begin{equation}
\label{dp}
 D(p)=\frac{1}{2} \Gamma(p) =\frac{1}{2}
\frac{p}{\langle p \rangle_0} y_{\alpha} G^2_F T^5,
\end{equation}
where $\langle p \rangle_0 \simeq 3.15 T$ is the average momentum
for a relativistic Fermi--Dirac distribution with zero chemical
potential, $y_e \simeq 4$, $y_{\mu, \, \tau} \simeq 2.9$ and $z_e
\simeq 0.1$, $z_{\mu,\,\tau} \simeq 0.04$.

The repopulation or refilling function
\begin{equation}
\label{ralpha}
R_{\alpha}(p) \simeq \Gamma(p) \left\{ K_{\alpha} -
\frac{1}{2}[P_0(p) + P_z(p)] \right\},
\end{equation}
with
\begin{equation}
\label{ratio} K_{\alpha} \equiv
\frac{N^{\text{eq}}(p,\mu)}{N^{\text{eq}}(p,0)},
\end{equation}
is determined from assuming thermal equilibrium for all species in
the background plasma, and that the $\nu_{\alpha}$ distribution is
also approximately thermal \cite{bvw}.  Physically, Eq.\
(\ref{ralpha}) means that as active neutrinos of some momentum are
converted to sterile neutrinos, a gap is created in the
Fermi--Dirac distribution.  The weak interaction processes
repopulate this depleted state, driving the ensemble back towards
equilibrium.

A separate but equivalent set of expressions parameterises the
evolution of the $\overline{\nu}_{\alpha} \leftrightarrow
\overline{\nu}_s$ system.  These are distinguished from their
ordinary counterparts with an overhead bar.

\section{Standard static/adiabatic limit: a brief outline}

Together with the requirement of $\alpha + s$ lepton number
conservation, we may derive from the QKEs an exact evolution
equation for the neutrino asymmetry $L_{\alpha}$, that is
\cite{bvw},
\begin{equation}
\label{exactdldt} \frac{dL_{\alpha}}{dt} = \frac{1}{2n_{\gamma}}
\int \beta [P_y(p) - \overline{P}_y(p)] N^{\text{eq}}(p,0) dp,
\end{equation}
where  the quantities $P_y$ and $\overline{P}_y$ are obtained from
numerically integrating the respective QKEs for the neutrino and
antineutrino systems given by Eq.\ (\ref{qkes}). The role of the
static/adiabatic limit approximation \cite{longpaper,bvw,v&w},
therefore, is to generate approximate expressions for $P_y$ that
are dynamically decoupled from the other variables in order to
minimise the computational effort.

The first step in the formal adiabatic procedure of Ref.\ \cite{bvw}
consists
of  setting the repopulation function to zero, i.e.,
\begin{equation}
\label{zeroR} R_{\alpha} \simeq 0,
\end{equation}
 thereby reducing the four-component QKEs to a system of three
coupled homogeneous differential equations. Further manipulation
produces the approximate equality
\begin{equation}
\label{adiabaticpy}
P_y(t) \simeq -\frac{\beta D}{D^2 + \lambda^2}
P_z(t),
\end{equation}
in the limit $D, \, |\lambda| \gg |\beta|$, such that Eq.\
(\ref{exactdldt}) becomes\footnote{The denominator in Eq.\
(\ref{adiabaticpy}), and therefore Eq.\ (\ref{approxdldt}), should
properly contain an additive term  $\beta^2$ to ensure the
equations' validity even when the condition $D \gg |\beta|$ does
not hold (see companion paper Ref.\ \cite{v&w}).  In this work,
however, we shall assume that the said condition is always met,
and omit the $\beta^2$ term for simplicity.}
\begin{equation}
\label{approxdldt}
\frac{dL_{\alpha}}{dt} \simeq
\frac{1}{2n_{\gamma}} \int \beta^2
\left[\frac{\overline{D}(\overline{N}_{\alpha} -
\overline{N}_s)}{\overline{D}^2 + \overline{\lambda}^2} -
\frac{D(N_{\alpha} - N_s) }{D^2 + \lambda^2} \right]
 dp.
\end{equation}
Detailed discussions of the standard adiabatic procedure may be
found in the companion paper Ref.\ \cite{v&w}.

In numerical studies, an analogous expression describing sterile
neutrino production for each momentum state is employed for
tracking the quantity $N_s(p)$ in Eq.\ (\ref{approxdldt}), while
the $\nu_{\alpha}$ distribution function is taken to be
\begin{equation}
\label{instant} N_{\alpha}(p) \simeq N^{\text{eq}}(p,\mu).
\end{equation}
This is the so-called instantaneous repopulation approximation,
which assumes that a depleted momentum state is immediately
refilled from the background medium so that thermal equilibrium is
always maintained.

A  na\"{\i}ve interpretation of Eqs.\ (\ref{ralpha}) and
(\ref{instant}) suggests that instantaneous repopulation leads
directly to an identically zero refilling rate, which seems to
argue for consistency between the central assumptions contained in
Eqs.\ (\ref{zeroR}) and (\ref{instant}).  However, the
instantaneous repopulation limit is also related to taking the
collision rate to infinity.  Thus the right hand side of Eq.\
(\ref{ralpha}) is an {\it a priori} undetermined finite and
generally nonzero function \cite{p&r}, that is apparently at odds
with the assumption of a vanishing repopulation rate.

\section{Extended adiabatic approximation}
\label{eaa}

The exact QKEs in Eq.\ (\ref{qkes}) are fundamentally a system of
four coupled first order differential equations, of both the
homogeneous and inhomogeneous  varieties, that is best displayed
in matrix form,
\begin{equation}
\label{qkematrix} \frac{\partial}{\partial t}
   \left(\begin{array}{c} P_x \\ P_y \\ P_z \\ P_0 \end{array}\right)
  =
   \left(\begin{array}{cccc}
   -D & -\lambda & 0 & 0 \\ \lambda & -D & -\beta & 0 \\
   0 & \beta & -D & -D   \\ 0 & 0 & -D & -D
   \end{array}\right)
   \left(\begin{array}{c} P_x \\ P_y \\ P_z \\ P_0 \end{array}\right)
  +
   \left(\begin{array}{c}
   0 \\ 0 \\ 2DK_{\alpha} \\ 2DK_{\alpha}
   \end{array}\right) \equiv {\cal K}^{\text{rp}} {\bf P}^{\text{rp}} + {\bf
   A}.
\end{equation}
Dealing firstly with the homogeneous part of Eq.\
(\ref{qkematrix}), we introduce an instantaneous diagonal basis
onto which we map the vector ${\bf P}^{\text{rp}} \equiv ({\bf P},
P_0)$ from its original fixed basis via the transformation
\begin{equation}
\left( \begin{array}{c} Q^{\text{rp}}_1 \\
            Q^{\text{rp}}_2 \\
            Q^{\text{rp}}_3 \\
            Q^{\text{rp}}_4 \end{array} \right)
\equiv {\bf Q}^{\text{rp}} = {\cal U}_{\text{rp}} {\bf
P}^{\text{rp}},
\end{equation}
where, by definition,
\begin{equation}{\cal K}_d^{\text{rp}} \equiv \text{diag}
\left(\Lambda_1,\, \Lambda_2,\, \Lambda_3, \, \Lambda_4 \right) =
{\cal U}_{\text{rp}}{\cal K}^{\text{rp}}{\cal U}^{-1}_{\text{rp}},
\end{equation}
such that the matrix ${\cal K}^{\text{rp}}_d$ is diagonal and
similar to ${\cal K}^{\text{rp}}$.

The eigenvalues $\Lambda_i$ are solutions to the quartic
characteristic equation
\begin{equation}
\Lambda^4+4D\Lambda^3+(5D^2+\lambda^2+\beta^2)\Lambda^2
+2D(D^2+\lambda^2+\beta^2)\Lambda+\beta^2D^2 = 0,
  \label{eq:CE2}
\end{equation}
formally given by
\begin{equation}
\Lambda_i = -D \pm \frac{\sqrt{D^2-\lambda^2-\beta^2
 \pm \sqrt{D^4+2D^2\lambda^2+\lambda^4 +2\beta^2\lambda^2-
 2\beta^2D^2+\beta^4}}}
{\sqrt{2}}.
   \label{eq:eval4}
\end{equation}
We are primarily interested in the limit $D,\, |\lambda| \gg
|\beta|$, in which case the eigenvalues are well approximated by
the leading order terms in the small $\beta$ power series
expansion of Eq.\ (\ref{eq:eval4}),
\begin{eqnarray}
\Lambda_1 & = & \Lambda_2^* =  -D+i\lambda+\frac{i}{2}
\frac{\beta^2\lambda}{D^2+\lambda^2}
       + {\cal O}(\beta^4),  \nonumber \\
\Lambda_3 & = & -\frac{1}{2}\frac{\beta^2 D}{D^2+\lambda^2}
                 + {\cal O}(\beta^4), \nonumber \\
\Lambda_4 & = & -2D + \frac{1}{2}\frac{\beta^2 D}{D^2+\lambda^2}+
{\cal O}(\beta^4).
   \label{eq:4evalbeta}
\end{eqnarray}
The transformation matrix ${\cal U}^{-1}_{\text{rp}}$ comprises
the column eigenvectors $\kappa_i^{\text{rp}}$, which for $\lambda
\neq 0$ are exactly
\begin{equation}
\label{u-1}
\kappa_i^{\text{rp}} = \left( \begin{array}{c}
   1 \\ -\frac{D+\Lambda_i}{\lambda} \\
   -\frac{\beta(D+\Lambda_i)^2}{\lambda \Lambda_i(2D+\Lambda_i)} \\
   \frac{\beta D(D+\Lambda_i)}{\lambda \Lambda_i(2D+\Lambda_i)}
   \end{array}\right),
\end{equation}
while its inverse ${\cal U}_{\text{rp}}$ consists of the row
vectors
\begin{eqnarray}
v_i^{\text{rp}} &&=   \Lambda_i (2D + \Lambda_i)\left(
\frac{1}{D^2} \prod_{j \neq i} \frac{D+\Lambda_j}{\Lambda_i -
\Lambda_j}, \quad -\lambda \prod_{j \neq i}
\frac{1}{\Lambda_i-\Lambda_j} , \right. \nonumber
\\
            && \left. \frac{\lambda }{\beta D^2}
            \prod_{j \neq i} \frac{1}{\Lambda_i-\Lambda_j}
                       \left[4 D^3 + 2D^2 (\Lambda_l + \Lambda_m + \Lambda_n) +
            D(\Lambda_l \Lambda_m + \Lambda_m \Lambda_n + \Lambda_n \Lambda_l)
            + \Lambda_l \Lambda_m \Lambda_n \right],
            \right.\nonumber \\
            && \left. \frac{\lambda}{\beta D}
\prod_{j \neq i} \frac{1}{\Lambda_i-\Lambda_j}
                       \left[4 D^2 + 2D (\Lambda_l + \Lambda_m + \Lambda_n) +
             \Lambda_l \Lambda_m + \Lambda_m \Lambda_n
+ \Lambda_n \Lambda_l \right] \right),
   \label{eq:U}
\end{eqnarray}
where the indices $l$, $m$ and $n$ are the three integers not
equal to $i$, e.g., for $i=1$, $l$, $m$ and $n$ are $2$, $3$ and
$4$ respectively.

Note that Eqs.\ (\ref{u-1}) and (\ref{eq:U}) do not apply if two
or more eigenvalues are degenerate, in which case the matrix
${\cal K}^{\text{rp}}$ may genuinely have less than four  distinct
eigenvectors thereby rendering ${\cal U}^{-1}_{\text{rp}}$
momentarily uninvertible, or, if four linearly independent
eigenvectors exist, alternative methods are needed for their
evaluation.  However, such circumstances are hard to come by and
have practically no influence on the outcome of this work. Thus
the exact QKEs [Eq.\ (\ref{qkematrix})] are equivalently
\begin{equation}
\label{qqkes}
\frac{\partial {\bf Q}^{\text{rp}}}{\partial t}
   = {\cal K}_d^{\text{rp}}{\bf Q}^{\text{rp}} + {\cal U}_{\text{rp}}{\bf A}
     - {\cal U}_{\text{rp}}\frac{\partial
{\cal U}_{\text{rp}}}{\partial t}^{-1}
     {\bf Q}^{\text{rp}},
\end{equation}
written in the new instantaneous basis where the transformation
matrices are valid.

The term ${\cal U}_{\text{rp}}\frac{\partial {\cal
U}_{\text{rp}}}{\partial t}^{-1}$ in Eq.\ (\ref{qqkes}) contains
explicit dependence on the derivatives of the parameters $D$,
$\lambda$ and $\beta$.  These are discarded in the adiabatic limit
(see Ref.\ \cite{bvw} for the relevant constraints\footnote{The
bounds in Ref.\ \cite{bvw} are calculated assuming $\frac{\partial
P_0}{\partial t} \simeq 0$, and should naturally be different from
those one would obtain with the incorporation of a finite
repopulation function. However, we expect the difference to be
minimal given the similarity between the relevant entries in the
two cases' respective transformation matrices in the appropriate
limit.}) so that the remaining terms form a set of four decoupled
inhomogeneous differential equations given by
\begin{equation}
\frac{\partial {\bf Q}^{\text{rp}}}{\partial t} \simeq {\cal
K}_d^{\text{rp}}{\bf Q}^{\text{rp}} + {\bf B},
  \label{eq:4matrixb}
\end{equation}
where ${\bf B} \equiv {\cal U}_{\text{rp}}{\bf A}$, or
equivalently in index form,
\begin{equation}
\label{decoupled4Q} \frac{\partial}{\partial t} Q^{\text{rp}}_i(t)
 \simeq \Lambda_i(t)Q_i^{\text{rp}}(t) + B_i(t), \;\; i =
 1,2,\ldots,4.
\end{equation}
Equation (\ref{decoupled4Q}) may be formally solved to give
\begin{equation}
Q_i^{\text{rp}}(t) =  e^{\int_0^t \Lambda_i(t')dt'}
Q_i^{\text{rp}}(0)+e^{\int^t \Lambda_i(t')dt'} {\cal I}_i,
\end{equation}
with the inhomogeneous segment of the QKEs contained entirely in
the integral
\begin{equation}
\label{intfactor} {\cal I}_i = \int_0^t e^{-\int^{t'}
\Lambda_i(t'')dt''}B_i(t') dt',
\end{equation}
thus amounting to the following solution for the vector ${\bf
P}^{\text{rp}}$,
\begin{eqnarray}
  \left(\begin{array}{c} P_x(t) \\ P_y(t) \\ P_z(t) \\ P_0(t)
  \end{array}\right)
  & = &
  {\cal U}_{\text{rp}}^{-1}(t) \left[
  \left(\begin{array}{cccc}
  e^{\int_0^t \Lambda_1(t')dt'} & 0 & 0 & 0 \\
  0 & e^{\int_0^t \Lambda_2(t')dt'} & 0 & 0 \\
  0 & 0 & e^{\int_0^t \Lambda_3(t')dt'} & 0 \\
  0 & 0 & 0 & e^{\int_0^t \Lambda_4(t')dt'} \\
  \end{array}\right)
   \left(\begin{array}{c} Q^{\text{rp}}_1(0) \\ Q^{\text{rp}}_2(0) \\
   Q^{\text{rp}}_3(0) \\ Q^{\text{rp}}_4(0)
  \end{array}\right)
  \right. \nonumber \\
  &  & \left. +
  \left(\begin{array}{cccc}
  e^{\int^t \Lambda_1(t')dt'} & 0 & 0 & 0 \\
  0 & e^{\int^t \Lambda_2(t')dt'} & 0 & 0 \\
  0 & 0 & e^{\int^t \Lambda_3(t')dt'} & 0 \\
  0 & 0 & 0 & e^{\int^t \Lambda_4(t')dt'} \\
  \end{array}\right)
  \left(\begin{array}{c} {\cal I}_1 \\ {\cal I}_2 \\ {\cal I}_3 \\ {\cal I}_4
  \end{array}\right) \right],
    \label{eq:4P}
\end{eqnarray}
 in the adiabatic limit.

  Observe in Eq.\
(\ref{eq:4evalbeta}) that the real components of the eigenvalues
$\Lambda_1$, $\Lambda_2$ and $\Lambda_4$ are of the order $D$,
while $\Lambda_3$ is proportional to $\beta^2$. This implies that
the exponentials $\exp[\int^t_0 \Lambda_i(t')dt']$, where
$i=1,2,4$, are rapidly damped relative to the ``decay'' time scale
of their $\Lambda_3$ counterpart in the homogeneous part of Eq.\
(\ref{eq:4P}) (for a full discussion, see Ref.\ \cite{v&w}). We
may therefore implement in Eq.\ (\ref{eq:4P}) the  collision
dominance approximation
\begin{equation}
e^{\int^t_0 \Lambda_i(t')dt'}  \to 0, \;\; i = 1,2,4,
   \label{eq:expk}
\end{equation}
 to obtain
\begin{eqnarray}
P_y(t) & \simeq & -\frac{D+\Lambda_3}{\lambda} e^{\int^t_0
\Lambda_3(t') dt'}Q^{\text{rp}}_3(0)
       - \sum_{i=1}^{4}\frac{D+\Lambda_i}{\lambda}
        e^{\int^t \Lambda_i(t')dt'}{\cal I}_i,
   \nonumber \\
P_z(t) & \simeq & -\frac{\beta(D+\Lambda_3)^2}{\lambda
\Lambda_3(2D+\Lambda_3)}e^{\int^t_0 \Lambda_3(t')
dt'}Q^{\text{rp}}_3(0)
       - \sum_{i=1}^{4}
       \frac{\beta(D+\Lambda_i)^2}{\lambda \Lambda_i(2D+\Lambda_i)}
       e^{\int^t \Lambda_i(t')dt'}{\cal I}_i,
    \nonumber \\
P_0(t) & \simeq & \frac{\beta D(D+\Lambda_3)}{\lambda
\Lambda_3(2D+\Lambda_3)} e^{\int^t_0 \Lambda_3(t') dt'}
Q^{\text{rp}}_3(0)
       + \sum_{i=1}^4
       \frac{\beta D(D+\Lambda_i)}{\lambda \Lambda_i(2D+\Lambda_i)}
        e^{\int^t \Lambda_i(t')dt'}{\cal I}_i,
  \label{eq:PyPz}
\end{eqnarray}
and consequently, from combining the above expressions for $P_y$
and $P_z$,
\begin{eqnarray}
P_y(t) =& &
\frac{\Lambda_3(2D+\Lambda_3)}{\beta(D+\Lambda_3)}P_z(t) \nonumber
\\
     && \qquad     + \sum_{j=1,2,4}\left(
    \frac{\Lambda_3(2D+\Lambda_3)}{\beta(D+\Lambda_3)}
    -\frac{\Lambda_j(2D+\Lambda_j)}{\beta(D+\Lambda_j)} \right)
    \frac{\beta(D+\Lambda_j)^2}{\lambda \Lambda_j(\Lambda_j+2D)}
    e^{\int^t \Lambda_j(t')dt'}{\cal I}_j.
   \label{eq:Py}
\end{eqnarray}
Note that the collision dominance approximation in Eq.\
(\ref{eq:expk}) is not immediately applicable to the indefinite
integrals associated with the inhomogeneous part of Eq.\
(\ref{eq:4P}) since the term ${\cal I}_i$ contains the factor
$\exp[-\int^t \Lambda_i(t')dt']$.

 With the quantity $B_i$ in Eq.\ (\ref{intfactor}) evaluating to
\begin{equation}
B_i = 2K_{\alpha} \frac{\lambda  \Lambda_i(2D+\Lambda_i)}
           {\beta D}
     \prod_{j \neq i} \frac{2D+\Lambda_j}{\Lambda_i - \Lambda_j},
   \label{eq:B}
\end{equation}
courtesy of Eq.\ (\ref{eq:U}), Eq.\ (\ref{eq:Py}) simplifies to
\begin{eqnarray}
P_y(t)  = && -\frac{\beta D}{D^2+\lambda^2}P_z(t)
  - e^{-\int^t (D - i \lambda) dt'}\int_{0}^{t}
  e^{\int^{t'} (D - i \lambda) dt''}\frac{\beta D K_{\alpha}}{D+i\lambda}
  dt' \nonumber \\
& &
  - e^{ -\int^t (D + i \lambda) dt'}\int_{0}^{t}
  e^{\int^{t'} (D + i \lambda) dt''}\frac{\beta D K_{\alpha}}{D-i\lambda}
  dt' \nonumber \\
&&  + \frac{4D}{\lambda}e^{-\int^t 2D dt'}\int_{0}^{t}
  e^{\int^{t'} 2D dt''}\frac{\beta \lambda D  K_{\alpha}}{D^2+\lambda^2}
  dt'
  + {\cal O}(\beta^3),
   \label{eq:Py2}
\end{eqnarray}
as a power series in the small expansion parameter $\beta$.
Clearly, the first term in the above expression is simply the
standard adiabatic result given by Eq.\ (\ref{adiabaticpy}).

The second item in Eq.\ (\ref{eq:Py2}) involving the inhomogeneous
parameter $K_{\alpha}$ may be partially solved using integration
by parts,
\begin{eqnarray}
\label{py2ndterm} \lefteqn{e^{-\int^t (D - i \lambda)
dt'}\int_{0}^{t}
  e^{\int^{t'} (D - i \lambda) dt''}\frac{\beta D K_{\alpha}}{D+i\lambda}
  dt' } \nonumber \\
& = &  e^{-\int^t (D - i \lambda) dt'}\int_{0}^{t} (D - i \lambda)
  e^{\int^{t'} (D - i \lambda) dt''}\frac{\beta D K_{\alpha}}{D^2+\lambda^2}
  dt' \nonumber \\
&=& \left. \frac{\beta D K_{\alpha}}{D^2 + \lambda^2}
\right|_{t'=t} - e^{-\int^t_0 (D - i\lambda) dt'} \left.
\frac{\beta D K_{\alpha}}{D^2 + \lambda^2} \right|_{t'=0}
\nonumber \\ && \quad- e^{-\int^t (D - i \lambda) dt'} \int^t_0
e^{\int^{t'} (D - i\lambda) dt''} \left[\frac{\beta D
}{D^2+\lambda^2} \frac{d K_{\alpha}}{dt'}  + K_{\alpha}
\frac{d}{dt'} \left(\frac{\beta D}{D^2 + \lambda^2}\right) \right]
dt'.
\end{eqnarray}
A sufficiently large damping parameter $D$ again allows the
approximation of $\exp [-\int^t_0 (D- i\lambda) dt'] \to 0$,
thereby promptly obliterating the second term on the right hand
side of the last equality.  The quantity  $\frac{d}{dt}
\left(\frac{\beta D}{D^2 + \lambda^2}\right)$ in the integral
turns out to be one of the elements constituting the $4\times 4$
matrix ${\cal U}_{\text{rp}} \frac{\partial {\cal
U}_{\text{rp}}}{\partial t}^{-1}$.  Consistency with the adiabatic
approximation then calls for the setting of $\frac{d}{dt}
\left(\frac{\beta D}{D^2 + \lambda^2}\right) \simeq 0$ such that
solutions to ${\bf P}_{\text{rp}}$ have no explicit dependence on
the time derivatives of the parameters $D$, $\lambda$ and $\beta$.

Further exploitation of the powerful technique of integration by
parts on Eq.\ (\ref{py2ndterm}) in conjunction
 with the aforementioned approximations thus yields
\begin{eqnarray}
\label{py2ndtermb} \lefteqn{e^{-\int^t (D - i \lambda)
dt'}\int_{0}^{t}
  e^{\int^{t'} (D - i \lambda) dt''}\frac{\beta D K_{\alpha}}{D+i\lambda}
  dt' } \nonumber \\
  & \simeq &  \left. \frac{\beta D K_{\alpha}}{D^2 + \lambda^2}
\right|_{t'=t} + \left[\sum_{n=1}^N
\left(\frac{-1}{D-i\lambda}\right)^n \frac{\beta D}{D^2 +
\lambda^2} \frac{d^{(n)}} {dt'^{(n)}} K_{\alpha} \right]_{t'=t}
\nonumber
\\ && \qquad - e^{-\int^t (D-i\lambda) dt'} \int^t_0 e^{\int^{t'}
(D-i \lambda) dt''} \left(\frac{-1}{D-i\lambda}\right)^N
\frac{\beta D}{D^2 + \lambda^2}
\frac{d^{(N+1)}}{dt'^{(N+1)}}K_{\alpha} dt'.
\end{eqnarray}
Above the neutrino decoupling temperature, the function
$K_{\alpha}$ is well approximated by
\begin{equation}
K_{\alpha} \simeq 1 + \frac{12 \zeta(3)}{\pi^2} \frac{e^{p/T}}{1 +
e^{p/T}} L_{\alpha},
\end{equation}
for small $L_{\alpha}$, and consequently,
\begin{equation}
\frac{d^{(n)}}{d t^{(n)}}K_{\alpha} \simeq \frac{12 \zeta
(3)}{\pi^2} \frac{e^{p/T}}{1 + e^{p/T}}\frac{d^{(n)}}{dt^{(n)}}
L_{\alpha},
\end{equation}
with the recognition that the dimensionless factor $e^{p/T}/(1 +
e^{p/T})$ is independent of time.  Since the quantity $\frac{d
L_{\alpha}}{dt}$ is of the order of $\beta^2$, the first term in
the sum containing this is equivalently an ${\cal O}(\beta^3)$
quantity that is generally negligible. The remainder of the sum
may be similarly ignored as higher order time derivatives of
$L_{\alpha}$ are successively smaller by factors of $\beta^2$,
leading to further simplification of Eq.\ (\ref{py2ndtermb}) to
\begin{eqnarray}
\lefteqn{e^{-\int^t (D - i \lambda) dt'}\int_{0}^{t}
  e^{\int^{t'} (D - i \lambda) dt''}\frac{\beta D K_{\alpha}}{D+i\lambda}
  dt'} \nonumber \\
 &\simeq & \left. \frac{\beta D K_{\alpha}}{D^2 + \lambda^2}
\right|_{t'=t} + {\cal O} (\beta^3),
\end{eqnarray}
and likewise for its complex conjugate in Eq.\ (\ref{eq:Py2}).

 The
last real term in the Eq.\ (\ref{eq:Py2}) may be shown to reduce
to
\begin{eqnarray}
\label{py4thterm} \lefteqn{\frac{4D}{\lambda}e^{-\int^t 2D
dt'}\int_{0}^{t}
  e^{\int^{t'} 2D dt''}\frac{\beta \lambda D  K_{\alpha}}{D^2+\lambda^2}
  dt'} \nonumber \\
&\simeq& \left. \frac{2 \beta D K_{\alpha}}{D^2 + \lambda^2}
\right|_{t'=t} \nonumber + \left[\sum_{n=1}^N
\left(\frac{-1}{2D}\right)^{n} \frac{2 \beta D}{D^2 + \lambda^2}
\frac{d^{(n)}} {dt'^{(n)}} K_{\alpha} \right]_{t'=t} \nonumber \\
&& \qquad- \frac{2D}{\lambda}e^{-\int^t 2D dt'} \int^t_0
e^{\int^{t'} 2D dt''} \left(\frac{-1}{2D}\right)^N \frac{\beta
\lambda}{D^2 + \lambda^2} \frac{d^{(N+1)}}{dt'^{(N+1)}}K_{\alpha}
dt' \nonumber \\ &\simeq& \left. \frac{2 \beta D K_{\alpha}}{D^2 +
\lambda^2} \right|_{t'=t} + {\cal O}(\beta^3),
\end{eqnarray}
by similar arguments such that Eqs.\ (\ref{py2ndtermb}) and
(\ref{py4thterm}) combine to produce
\begin{equation}
\label{repoppy}
P_y(t) \simeq - \frac{\beta D}{D^2 + \lambda^2}
P_z(t) + {\cal O}(\beta^3).
\end{equation}
Equation (\ref{repoppy}) clearly shows the miraculous cancellation
between terms pertaining to a nonzero repopulation function
originally present in Eq.\ (\ref{eq:Py2}) to first order in
$\beta$. The final expression for $P_y(t)$ is therefore identical
to the standard adiabatic result given by Eq.\ (\ref{adiabaticpy})
in the limit of interest.

Our next task is to verify that the assumption of instantaneous
repopulation in Eq.\ (\ref{instant}) is indeed correct.  For this
purpose, we begin by expressing $P_z$ as a function of $P_0$ by
way of Eq.\ (\ref{eq:PyPz}),
\begin{equation}
\label{pzp0} P_z(t) = - \left(1+ \frac{\Lambda_3}{D} \right)
P_0(t) + \sum_{j=1,2,4}
 \frac{\beta(D+\Lambda_j)(\Lambda_3-\Lambda_j)}
{\lambda \Lambda_j (2D+\Lambda_j)}
e^{\int^t \Lambda_j(t') dt'} {\cal I}_i.
\end{equation}
To the lowest order in $\beta$, Eq.\ (\ref{pzp0}) is equivalently
\begin{equation}
P_z(t) = - P_0(t) + \frac{4(D^2 + \lambda^2)}{\beta \lambda} e^{-
\int^t 2D dt'} \int^t_0 e^{\int^{t'} 2D dt''} \frac{\beta \lambda
D K_{\alpha}}{D^2 + \lambda^2} dt'+ {\cal O}(\beta^2),
\end{equation}
on which we apply the same technique of integration by parts in
conjunction with the rationale employed  previously to eliminate
the various time derivatives to obtain
\begin{eqnarray}
\label{p0pz} P_z(t) &\simeq&  - P_0(t) + 2 K_{\alpha}(t) +
2\left[\sum_{n=1}^N \left(\frac{-1}{2D}\right)^{n} \frac{d^{(n)}}
{dt'^{(n)}} K_{\alpha} \right]_{t'=t} \nonumber \\ && \qquad -
\frac{2(D^2 + \lambda^2)}{\beta \lambda} e^{-\int^t 2D dt'}
\int^t_0 e^{\int^{t'} 2D dt''} \left(\frac{-1}{2D}\right)^N
\frac{\beta \lambda}{D^2 + \lambda^2}
\frac{d^{(N+1)}}{dt'^{(N+1)}}K_{\alpha} dt' \nonumber \\ &\simeq &
- P_0(t) + 2 K_{\alpha}(t) + {\cal O}(\beta^2).
\end{eqnarray}
The last approximate equality in Eq.\ (\ref{p0pz}) is identically
\begin{equation}
\frac{N_{\alpha}(p)}{N^{\text{eq}}(p,0)} \simeq K_{\alpha} \equiv
\frac{N^{\text{eq}}(p,\mu)}{N^{\text{eq}}(p,0)},
\end{equation}
which clearly affirms the validity of the instantaneous
repopulation assumption to the lowest order in $\beta$.

\section{Conclusion}

We have taken the exact four-component QKEs for a two-flavour
active--sterile neutrino system and derived from them an
approximate evolution equation for the relic neutrino asymmetry in
which the role of repopulation is methodically embedded. The
consequence of including a finite refilling function is to
generate higher order terms which are readily discarded in the
high temperature epoch of interest, thereby yielding a rate
equation identical to that found earlier in the standard adiabatic
limit where the said function is taken to be negligible.  The
formal procedure developed in the present work to establish this
result has been labelled the extended adiabatic approximation.

Frequently adopted in numerical studies with notable accuracy, the
assumption of instantaneous repopulation is also shown to arise
naturally in the extended adiabatic limit. We have thus furnished
a rigorous justification for the superficially
incompatible assumptions of a negligible refilling function and
instantaneous repopulation in the regime where collisions are the
dominant asymmetry  generation mode.

\acknowledgments{This work was supported in part by the Australian
Research Council and in part by the Commonwealth of Australia's
postgraduate award scheme.}


\begin{thebibliography}{99}

\bibitem{longpaper}
R. Foot and R. R. Volkas, Phys.\ Rev.\ D {\bf 55}, 5147 (1997).

\bibitem{qke}
B. H. J. McKellar and M. J. Thomson, Phys.\ Rev.\ D {\bf 49}, 2710
(1994)

\bibitem{qke2}
For foundational work see A. Dolgov, Sov.\ J.\ Nucl.\ Phys.\ {\bf
33}, 700 (1981); R. A. Harris and L. Stodolsky, Phys.\ Lett.\ {\bf
116B}, 464 (1982); L. Stodolsky, Phys.\ Rev.\ D {\bf 36}, 2273
(1987); G. Raffelt, G. Sigl and L. Stodolsky, Phys.\ Rev.\  Lett
{\bf 70}, 2363 (1993); M. J. Thomson, Phys.\ Rev.\ A {\bf 45},
2243 (1992).


\bibitem{bvw}
N. F. Bell, R. R. Volkas and Y. Y. Y. Wong, Phys.\ Rev.\ D {\bf
59}, 113001 (1999).


\bibitem{v&w}
R. R. Volkas and Y. Y. Y. Wong, companion paper,
hep-ph/0007185.

\bibitem{p&r}
P. Di Bari and R. Foot,
Phys.\ Rev.\ D {\bf 61}, 105012 (2000).

\bibitem{rn}
D. N\"{o}tzold and G. Raffelt, Nucl.\ Phys.\  {\bf B307}, 924
(1988).





\end{thebibliography}
\end{document}